\documentclass[%
 aip,
 amsmath,amssymb,
 reprint,%
]{revtex4-1}

\usepackage{graphicx}
\usepackage{dcolumn}
\usepackage{bm}

\usepackage[utf8]{inputenc}
\usepackage[T1]{fontenc}
\usepackage{mathptmx}
\usepackage{xfrac}
\usepackage{gensymb}
\usepackage{amsmath}
\usepackage{mathtools}
\usepackage[labelformat=simple]{subcaption}

\usepackage{xcolor}
\usepackage[normalem]{ulem}
\usepackage{placeins}
  
\newcommand\hl{%
  \bgroup
  \expandafter\def\csname sout\space\endcsname{\bgroup \ULdepth =-.8ex \ULset}%
  \markoverwith{\textcolor{green}{\rule[-.5ex]{.1pt}{2.5ex}}}%
  \ULon}

\begin{document}

\preprint{AIP/123-QED}

\title[]{On Free Moving Micron-Sized Droplet-Particle Collisions}

\author{Tushar Srivastava}
\affiliation{%
School of Chemical and Process Engineering, University of Leeds, Leeds, UK
}%

\author{Karrar H.Al-Dirawi}
\affiliation{%
School of Natural Sciences, Department of Mathematics, University of Manchester, Manchester, UK}%

\author{Benjamin Lobel}
\affiliation{%
School of Mathematics, Statistics, Chemistry and Physics, Murdoch University, Australia}%

\author{Andrew E. Bayly}
\email{a.e.bayly@leeds.ac.uk}
\affiliation{%
School of Chemical and Process Engineering, University of Leeds, Leeds, UK
}%

\date{\today}

\begin{abstract}
Predictive modelling of agglomeration in spray drying and particle capture in aerosol scavenging requires a fundamental understanding of droplet–particle collisions. The study complements prior work by investigating mid-air collisions between free micron-sized spherical droplets and particles with a size ratio of three. Particle wettability and density are varied to elucidate the mechanisms governing collision outcomes and the role of collision offset. Results show that particle density determines whether a particle is engulfed by the droplet or remains at the droplet interface during capture, while high wettability suppresses particle separation even in glancing collisions. A modified effective Weber number incorporating particle density and wettability is proposed to map collision outcomes. To assess its robustness, the present data are combined with literature results in a unified regime map. The regime boundaries separating collision outcomes collapse when the size ratio and Ohnesorge number are held constant. However, at a given collision offset, variations in size ratio and Ohnesorge number alter the critical effective Weber number for particle separation through changes in collision geometry and viscous resistance.
\end{abstract}

\maketitle



\section{Introduction}

Droplet-particle (D-P) collision is an ubiquitous phenomena encountered in a variety of applications. For example, coal dust reduction in coal-fired plants \cite{wu2016abatement}, agglomeration in food and dairy spray dryers \cite{frohlich2023nozzle}, synthesis of Metal Matrix Composites (MMCs) through spray atomization \cite{wu1992interaction}, aerosol scavenging and ice nucleation caused by cloud droplets and aerosol \cite{byrne1993scavenging,tinsley2000effects}, Fluid Catalytic Cracking (FCC) of atomized heavy oil droplets \cite{malgarinos2017numerical}, fluidised bed coating of powders \cite{Tenou}, and so on.

In general, the dynamics of droplet-particle (D-P) collisions are governed by a complex interplay of inertial, capillary, viscous, electrical, and other phoretic forces, with their relative importance dictated by the characteristic length scale of the system \cite{ardon2015laboratory}. For D-P systems with characteristic dimensions exceeding $1\ \mu\mathrm{m}$, inertial, capillary, and viscous forces dominate the interaction \cite{zhan2023droplet}. Understanding the interplay between these forces and the underlying mechanism governing D-P interaction provides the foundation for developing models and predicting collision outcomes \cite{speirs2023capture}.

In droplet–particle collisions, the droplet-to-particle diameter ratio $\Delta = D_d / D_p$ may be less than, equal to, or greater than unity. To date, head-on collisions with $\Delta \leq 1$ have received substantial attention, yielding important insights into post-collision droplet behaviour such as rebound \cite{bordbar2018maximum,yoon2022promoting,khojasteh2019review, geng2023collision,mitra2013droplet}, spreading \& deposition \cite{bakshi2007investigations,banitabaei2017droplet,khurana2019phenomenology,yang2018simulation}, and splashing \cite{hardalupas1999experimental,charalampous2017collisions,sykes2022droplet} on spherical particles . In these studies, the droplet dynamics are commonly described using two dimensionless numbers: the droplet Weber number, $We_d = \rho_d v_r^2 D_d / \sigma$, representing the ratio of inertial to capillary forces, and the Ohnesorge number, $Oh_d = \mu / \sqrt{\rho_d \sigma D_d}$, representing the relative importance of viscous forces compared to inertia and capillary forces. Here, $\rho_d$ is the droplet density, $v_r$ the relative velocity, $\sigma$ the surface tension, and $\mu$ the dynamic viscosity. For oblique collisions, the impact parameter, $B = 2b / (D_d + D_p)$, becomes important, where $b$ is the lateral displacement between the droplet and particle centers at impact. With varying $B$, when the droplet and particle are of comparable size, or when $\Delta > 1$, the droplet either adheres to the particle, leading to capture (often termed agglomeration or deposition), or detaches after the collision, resulting in separation \cite{dubrovsky1992particle,pawar2016experimental,le2024novel,fan2024numerical,islamova2022droplet}. Although $\Delta$ influences the force balance and the critical conditions that govern the collision outcomes, experimental studies across different $\Delta$ values remain limited. Therefore, the role of $\Delta$ in D–P collision dynamics, particularly for $\Delta > 1$, is still not fully understood.

Collision outcomes, including capture and separation, are often represented on a $B$–$We_d$ regime map \cite{le2024novel,pawar2016experimental,tkachenko2022experimental} to identify the boundary separating these outcomes. In the limiting case of $\Delta \ll 1$, $We_d$ captures the droplet dynamics inherent in the collision \cite{islamova2022droplet}. The reverse holds for $\Delta \gg 1$, where the particle Weber number, $We_p = \rho_p v_r^2 D_p / \sigma$, reflects the inertial response of the particle \cite{speirs2023capture}, with $\rho_p$ as the particle density. However, for intermediate values of $\Delta$, both droplet and particle inertia could contribute to D–P collision dynamics \cite{zhu2025dynamics}. The effect of particle wettability must also be considered while characterizing collision outcomes \cite{sechenyh2016experimental,islamova2022droplet}. For example, hydrophobic particles, with low wettability and high contact angles, tend to promote rebound due to weak adhesion and limited energy dissipation \cite{wu2021simulating,zhan2023droplet}, whereas hydrophilic particles enhance spreading and viscous dissipation, leading to greater energy loss and an increased probability of capture \cite{speirs2023capture}. Current studies examining the combined influence of particle density and wettability across different $B$ remain scarce, highlighting the need for further investigation. 

Unlike $We_d$, the role of $Oh_d$ in D–P collision outcomes has received limited attention \cite{yoon2021maximal}. Most experimental studies employ millimeter-sized droplets, which produce small $Oh_d$ values, rendering viscous effects negligible \cite{pawar2016experimental,sechenyh2016experimental}. For the same $\Delta$, however, smaller droplets exhibit higher $Oh_d$ and stronger viscous resistance \cite{zhan2023droplet}, which stabilizes the droplet against deformation and can suppress particle detachment. Recent experiments have explored sub-millimeter droplets \cite{le2024novel}, but these were carried out at fixed $Oh_d$, limiting understanding of its influence. Systematic studies that vary $Oh_d$ and enable comparison across different conditions are needed to generate robust data, support meaningful comparisons, and guide the development of predictive models for collision outcomes. This is important in particular for practical applications, where droplets and particles are often micron-sized \cite{zhu2025dynamics}.

The conventional approach for conducting D–P collision experiments involves holding either the droplet or the particle stationary while the other is released or propelled toward it \cite{mitra2015collision,speirs2023capture,islamova2022droplet}. However, the momentum transfer experienced by a droplet or particle upon impacting a stationary target differs from that occurring during a collision between two free-moving bodies \cite{yoon2022adaptive,wu2021simulating,zhu2025dynamics}. Collisions between free-moving droplets and particles also induce rotation, which can consume a portion of the initial kinetic energy and also influence droplet deformation, thereby affecting the overall collision outcome \cite{tkachenko2022experimental}.

\begin{figure*}
	\centering
		\includegraphics[scale=0.8]{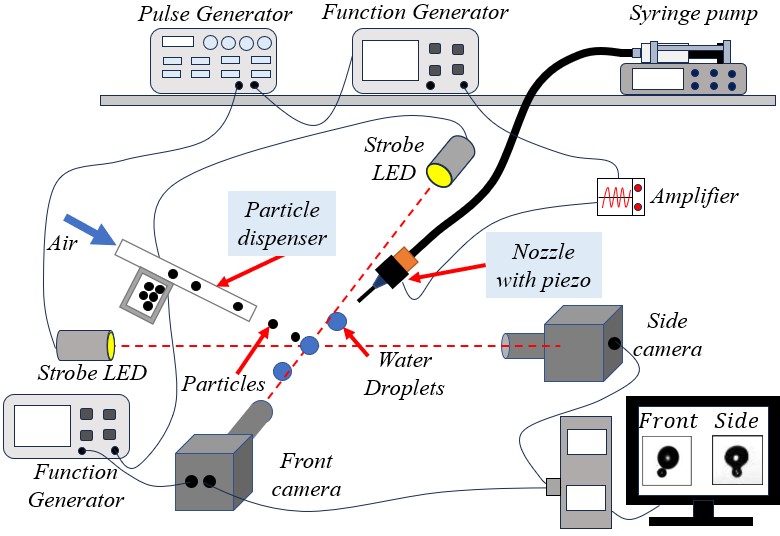}
	\caption{Schematic of the experimental setup employed in the present study}
	\label{exp_setup}
\end{figure*}

\begin{figure}
	\centering
		\includegraphics[scale=0.5]{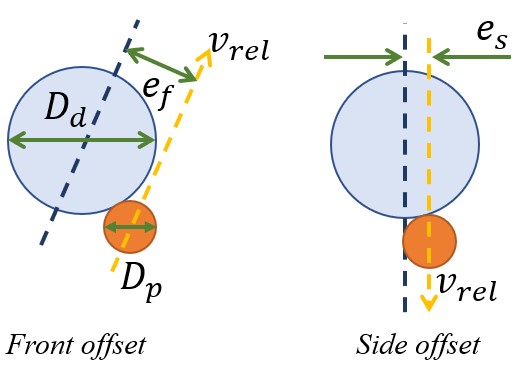}
	\caption{Schematic representation of the collision offset        as observed from front and side camera}
	\label{coll_offset}
\end{figure}

In this study, we present the first experimental investigation of the collision dynamics of a moving micron-sized droplet–particle system with $\Delta = 3$ and varying $B$. Particles with varying particle-to-droplet density ratio, $\lambda$, and wettability, $\theta_w$, are chosen for the investigation. Experiments are conducted using a custom-built setup equipped with two high-speed cameras, a droplet dispenser, and a particle dispenser with alignment controls to capture droplet–particle collisions in air. The high-speed images are analyzed to obtain mechanistic insights into the influence of particle density and wettability on collision outcomes. Furthermore, by combining a zero-momentum frame of reference analysis with a simple geometrical approach, we derive a new effective parameter, $We_{peffw}$, which accounts for both droplet and particle inertia in moving D–P collisions and use it to characterize and compare collision outcomes. Comparison of the present experimental results with literature data obtained at different $Oh_d$, $\lambda$, $\theta_w$, and $\Delta$ in $We_{peffw}$ vs $B$ space provides new insights into the role of these parameters in collision dynamics. The findings have the potential to make a significant contribution to the development of predictive models with practical relevance across a range of applications, including spray drying and aerosol scavenging.

\section{Methodology}
Fig. \ref{exp_setup} shows the schematic of the experimental setup used in the current work. The setup is a modification over the previous setup used for investigating droplet-droplet collisions \cite{al2019new}. During experiments, continuous fluid supplied by the syringe pump created a liquid jet when passed through a nozzle. At the same time, square wave signals programmed in a function generator were transmitted through a 20 $\times$ amplifier (PiezoDrive PDU150CL) to a piezo chip that is built into the nozzle. The piezo chip vibrated and excited the jet to create a reproducible droplet stream. Optimizing the frequency and amplitude of the wave signal led to the generation of mono-disperse droplets. The present work employed a 0.150 mm ID nozzle for the experiments.

\begin{table}[ht]
\centering
\caption{Table listing the particle properties chosen for the study.}
\label{particle_table}
\begin{tabular}{c c c c}
\hline
S. no. & Material name & Density, $\rho_p$ (kg/m$^3$) & Wettability, $\theta_w$ ($^\circ$) \\
\hline
1 & GB  & 2500 & 11 $\pm$ 1 \\
2 & PB  & 1000 & 90 $\pm$ 2 \\
3 & TGB & 2500 & 88 $\pm$ 2 \\
\hline
\end{tabular}
\end{table}

For releasing particles, a dispensing system was introduced. The system consisted of a capped reservoir linked to a tube connected to a compressed air supply. When perturbations or tapping were applied to the cap, the resulting disturbance caused the particles to ascend into the air stream and exit the tube. A controlled release of the compressed air regulated the dispersion of the particles upon exiting the tube. In addition, a sufficient gap was maintained between the particle dispensing system and the nozzle to ensure that the droplets generated from the nozzle were not affected by the air flow. Though the particle dispersion was controlled with the current particle dispensing system, it was difficult to ensure on-axis D-P collisions. Therefore, to capture the collision phenomenon, two high-speed cameras were used; a front camera (Photron mini AX100; master) and a side camera (Photron Fastcam SA5; slave) which recorded images at 20,000 fps while in-sync. The master camera was attached to a Navitar microscopic zoom lens. While recording the collision and its outcomes, the magnification was set as 4$\times$. The corresponding field of view (FOV) was 384 $\times$ 384 pixels, and the image resolution was 10 $\mu m/$pixel. Thus, the uncertainty in the measurement of the droplet and particle sizes was 5\%. Two strobe LEDs synchronized with the two high-speed cameras were used for illumination. The strobe lights provided 10 ns pulse duration which enabled more control over the exposure time. To facilitate droplet-particle (D-P) collisions, a XYZR micro-traverser attached to the base of the nozzle post was adjusted to align the droplet stream with the incoming particles.

To investigate the role of particle properties in influencing the outcomes of droplet–particle collisions, spherical glass beads (Sigmund Lindner GmbH) and polyethylene microspheres (Cospheric LLC) were used. A portion of the glass beads was further treated via silanisation \cite{lobel2021interparticle}, to modify their surface wettability. In the following discussion, we will use abbreviated terms for the three particle types: GB for glass beads, PB for polyethylene beads, and TGB for treated silanised glass beads. Table \ref{particle_table} lists the particles with different density and wettability values chosen for the investigation. The density values were taken from the manufacturer’s specifications. To characterize particle wettability, static contact angles, $\theta_w$, were measured using a tensiometer (Biolin Scientific) by depositing a water droplet onto a flat slide composed of the same material as the corresponding beads. For the treated glass beads, a glass slide was silanised using the same mol/m$^2$ of silane as used for the beads, ensuring that the resultant contact angle is the same.

\begin{figure*}[t]
	\centering
	\includegraphics[scale=0.5]{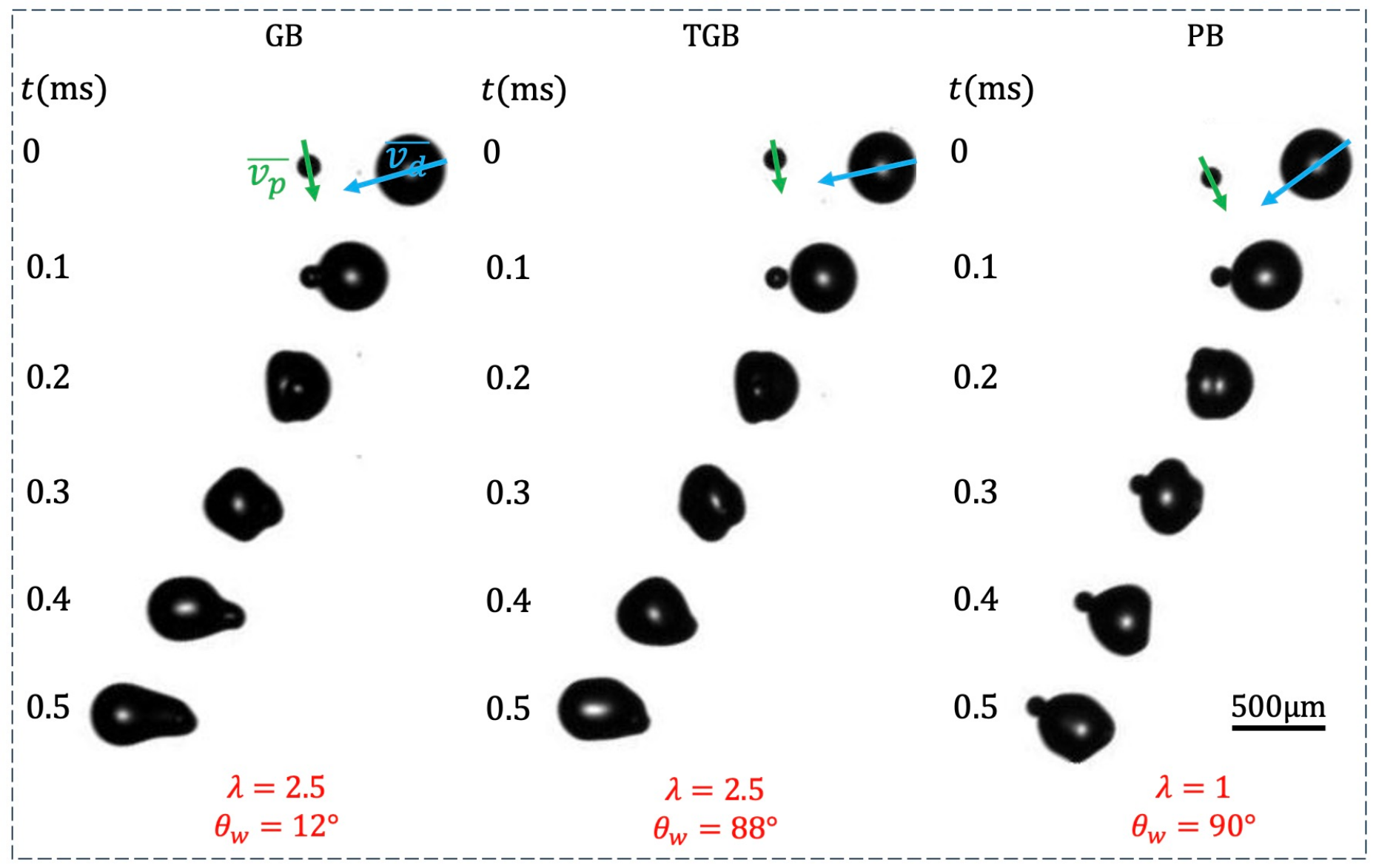}
	\caption{Time resolved snapshots showing D-P interactions for all three particles: GB, TGB, and PB at $We_d\approx67$ and $B\approx0.1$.}
	\label{DPlow}
\end{figure*}

To process the recorded images and identify the collision parameters, an image processing code was developed in MATLAB. The calibration of the images was conducted using a stage micrometer. The code extracted the droplet diameter, $D_d$, particle diameter, $D_p$, droplet velocity, $v_d$, particle velocity, $v_p$, relative velocity, $v_r$, and impact parameters as viewed from the front and side, $B_f$ and $B_s$ respectively. The impact parameters represent the dimensionless offset and are expressed as: $B_f = 2e_f/(D_d + D_p)$ and $B_s = 2e_s/(D_d + D_p)$. For simplicity and accuracy, only the recorded images exhibiting $B_s = 0$ were processed. Figure \ref{coll_offset} illustrates how $B_f$ was measured during image processing. In the subsequent text, we will use $B$ instead of $B_f$ to define the impact parameter or front offset.

During experiments, the air flow rate (facilitating particle release) and droplet flow rate were manipulated to modify the relative velocity ($v_r$) and therefore the Weber number ($We_d$ or $We_p$). The droplet diameter, $D_d$, was determined through image processing and was found to be 399.61 $\pm$ 13.12 $\mu$m. To ensure spherical and uniform droplet size, the major diameter ($D_{\text{major}}$), minor diameter ($D_{\text{minor}}$), and circularity were measured. The axis ratio error ($\frac{(D_{major}-D_{minor})}{D_{major}} \times 100$) was maintained at $\leq 8$\%, and the circularity was $\geq$ 0.95 for all droplets included in the analysis (\textcolor{blue}{refer to S1 in the Supplementary material}). Similarly, the particle diameter was determined as 132.89 ± 9.43 $\mu$m. Subsequently, the ratio of the droplet diameter to particle diameter ($D_d/D_p$) was evaluated and found to be 3.007 ± 0.232. 

The uncertainty in the measurement of $v_d$ and $v_p$ was determined by evaluating the mean of the positional data collected from at least five frames recorded before the collision. In addition to the impact parameter, the droplet Weber number, $We_d$, and the particle Weber number, were determined to quantify the outcomes of the collision. Since uncertainty in the measurement of $D_d$, $D_p$, $v_d$, and $v_p$ were known, we used the principle of error propagation to determine to determine uncertainty in the determination of $We_d$ and $We_p$. The error values were incorporated while constructing the regime maps to ascertain the critical boundary demarcating the collision outcomes.

\section{Result \& Discussions}
After processing the high-speed camera images, we analyzed each instantaneous image to study the stages of D-P interaction over time. Our goal was to determine how the particle's physical properties: the density and wettability, influenced the collision outcomes. 

Figure \ref{DPlow} presents time-resolved snapshots of a water droplet interacting with the particles: GB, TGB, and PB, at a similar droplet Weber number, $We_d \approx 67$, and a low collision offset, $B \approx 0.1$. As observed, the droplet–particle (D–P) interaction induces noticeable droplet deformation. In the early stage of the collision, capillary waves originate at the droplet–particle interface and propagate upward along the droplet periphery, contributing to its deformation (\textcolor{blue}{refer to S2 in the supplementary material}). During this process, the GB and TGB particles appear to penetrate the droplet, while the PB particle remains entrapped at the droplet interface.

The variation in collision outcomes resembles that observed for solid particles impacting planar liquid surfaces \cite{vella2015floating, chen2018entrapping}. Depending on the impact velocity and particle wettability, outcomes such as bouncing, interfacial entrapment, or full penetration may occur. In general, particles with a static contact angle $\theta_w \gtrsim 110\degree$ tend to bounce upon impact with a liquid surface \cite{chen2018entrapping}. In D-P collisions, the Laplace pressure arising from droplet curvature enhances the interfacial resistance to deformation and particle penetration \cite{mitra2015collision}, thereby promoting rebound. However, in the present experiments on D–P collisions with $\Delta = 3$, only particle entrapment or penetration was observed during head-on impacts for both hydrophilic ($\theta_w \approx 10\degree$) and hydrophobic ($\theta_w \approx 90\degree$) particles across the range of impact conditions considered in this study. These observations indicate that, at this size ratio and particle wettability values, the restoring effect of Laplace pressure is insufficient to induce particle rebound. The mechanisms governing the collision outcomes in Fig. \ref{DPlow} can be further understood through a simple scaling analysis.

Consider a D–P collision with a size ratio, $\Delta > 1$. The inertial forces acting on the particle promote its penetration into the droplet, while this motion is opposed by the capillary force, the pressure force (Laplace pressure), and the viscous stresses exerted by the droplet \cite{mitra2015collision}. The capillary force, $F_c$, depends on the contact angle formed by the particle at the three-phase contact line (TPCL) and can be expressed as $F_c = (\pi/2) D_p \sigma ( \cos( \theta_w + 2\beta) - \cos\theta_w )$ \cite{chen2018entrapping,lee2008impact,mitra2015collision}. Here, $\theta_w$ is the wetting angle of the particle, and $\beta$ is the penetration angle, which varies with the depth of particle immersion within the droplet. The maximum value of $F_c$ therefore scales as $F_{c,\max} \sim D_p \sigma \sin^2(\theta_w/2)$. In the limit $Oh \ll 1$, the viscous stresses exerted by the droplet on the particle can be neglected. For D–P collisions considered at the same size ratio $\Delta$, the relative contribution of the Laplace pressure force remains unchanged. Hence, the capillary-induced deceleration of the particle after its interaction with the droplet interface can be estimated as $a \sim \sfrac{\sigma \sin^2(\theta_w/2)}{\rho_p D_p^2}$. Here, $\rho_p$ is the density of the particle. Assuming that the time scale, $t_c$, associated with the penetration of the particle scales as capillary time scale, $t_c \sim (\rho_d D_p^3/\sigma)^{0.5}$, the critical velocity required for the particle to penetrate the droplet can be expressed as: $v_c \sim a t_c$. Substituting the values of $a$ and $t_c$ gives us:
\begin{equation}
    v_c  \sim \frac{sin^2(\theta_w/2)}{\lambda}\left\{\frac{\sigma}{\rho_d D_p}\right\}^{0.5}
    \label{Wed_depend}
\end{equation}
Equation \ref{Wed_depend} shows that the critical velocity, $v_c$, increases with decreasing particle-to-droplet density ratio, $\lambda = \rho_p / \rho_d$, and with increasing contact angle, $\theta_w$. Therefore, a PB particle, which has a large $\theta_w$ and a small $\lambda$, exhibits the highest $v_c$ and experiences the most pronounced deceleration upon impact compared with GB and TGB particles at similar $v_{rel}$ or $We_d$. This behaviour explains the observed entrapment of the PB particle at the droplet interface. 

Although the TGB and PB particles have similar wettability, the larger $\lambda$ of the TGB particle lowers the $v_c$ required for penetration. Thus, at $We_d \approx 67$, the TGB particle penetrates the droplet, whereas at lower $We_d$ it becomes entrapped at the interface (\textcolor{blue}{refer to S3 in the Supplementary Material}). The GB particle, with its higher $\lambda$ and smaller $\theta_w$, requires the lowest $v_c$ for penetration and experiences the least deceleration among the three particle types. That is why we recorded complete wetting of the GB particle by the droplet over the entire range of $We_d$ examined in this study.

\begin{figure*}
	\centering
		\includegraphics[scale=0.5]{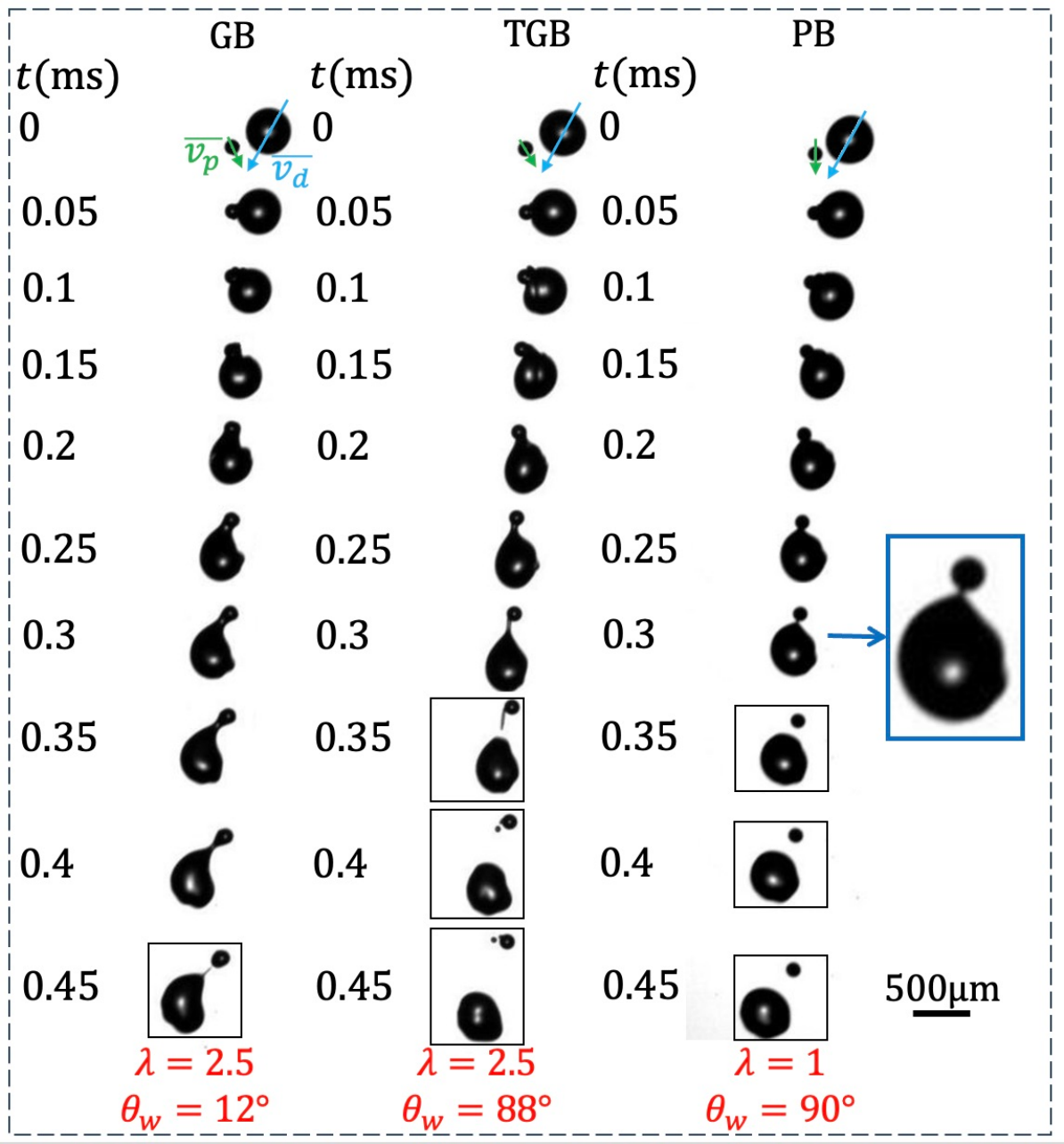}
	\caption{Time resolved snapshots showing D-P interactions for all three particles: GB, TGB, and PB at $We_d \approx 65$ and $B=0.7$.}
	\label{separation}
\end{figure*}

\begin{figure*}[ht]
	\centering
		\includegraphics[scale=0.70]{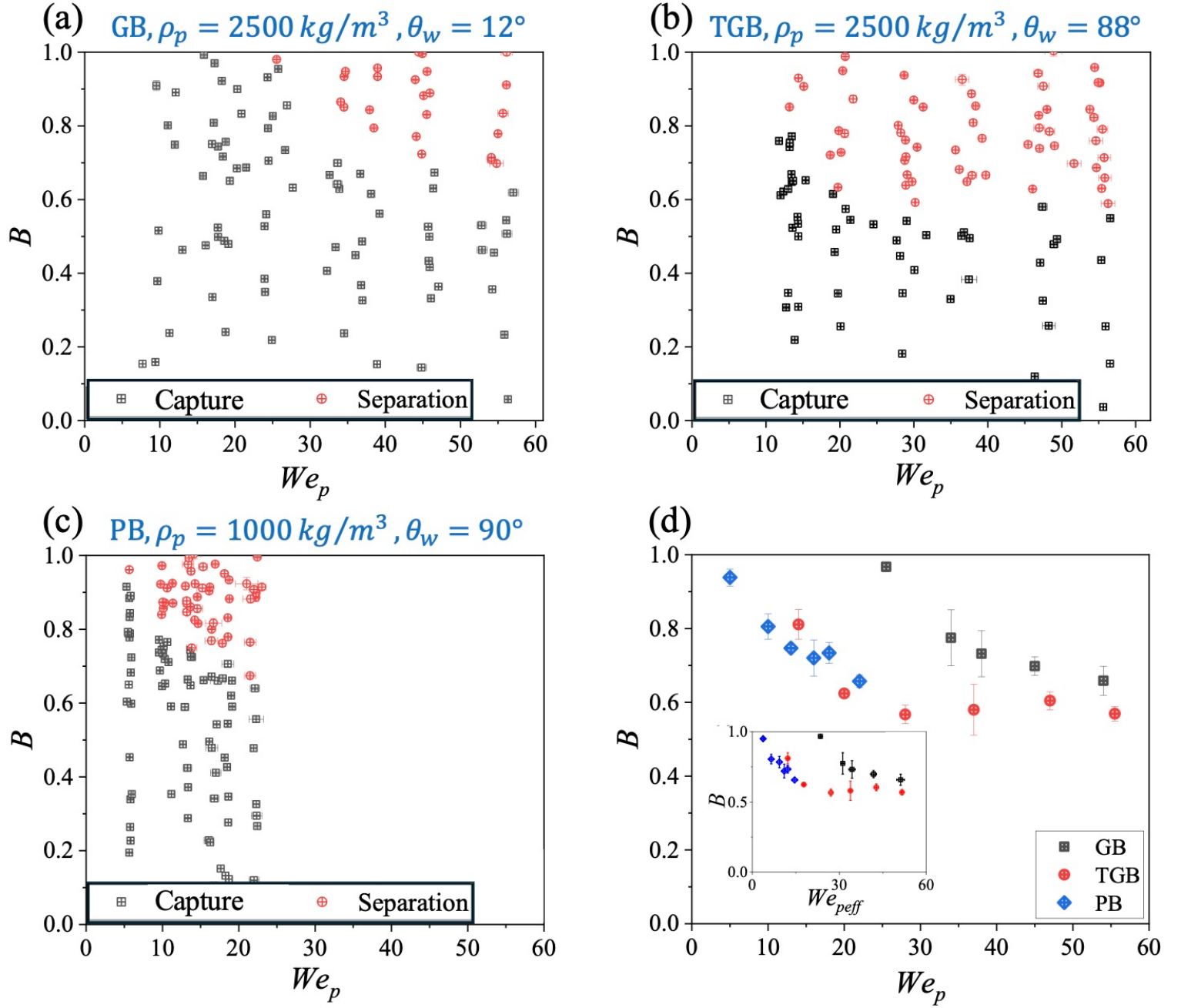}
	\caption{Plots showing coalescence (black symbols) and separation (red symbols) outcomes for beads: Glass beads (GB), Polyethylene beads (PB), and Treated Glass beads (TGB), colliding with water droplet at different $We_p $ and $B$ with size ratio, $\Delta=3$.}
	\label{reg_map}
\end{figure*}

Increasing $\lambda$, and hence the particle inertia, also enhances droplet elongation after collision. This behaviour is evident in Fig. \ref{DPlow}, where the D–P interaction with GB leads to minimal particle deceleration and the largest droplet deformation, followed by TGB and PB, which shows the least. A similar trend is observed for D–P collisions at higher $B$. Time-resolved images in Fig. \ref{separation} illustrate D–P interactions of the three particles, leading to separation at $We_d \approx 65$ and $B \approx 0.7$. After impact, the droplet and particle merge and rotate as a single body. A liquid ligament then forms between them and continues to stretch until the particle detaches from the droplet. The process resembles the stretch–separation mode reported for droplet–droplet collisions at moderate to high $B$ \cite{al2021inertial}.

The GB and TGB particles, which have larger values of $\lambda$, generate longer and more elongated ligaments than the PB particle, which produces only a short ligament (see inset of Fig. \ref{separation}). This behaviour arises because an increase in $\lambda$, and hence in particle inertia, enhances the transfer of momentum and kinetic energy to the droplet during the collision, leading to ligament extension. Since $We_d \lambda$ reduces to $\sim We_p / \Delta$, the particle Weber number $We_p$ serves as a more appropriate parameter for comparing collision outcomes at fixed $\Delta$.

We also observe that, like PB, the ligament is thinner for TGB compared to GB. This reduced ligament thickness can be attributed to the large contact angles formed by TGB and PB particles formed at the TPCL. As a result, the ligament exhibits a steeper curvature and is thinner than that of GB, where the particle forms a smaller contact angle. A thicker ligament stretches further before breakup, which explains the delayed separation of the GB particle from the droplet observed in Fig. \ref{separation}.

Apart from interfacial effects, rotational motion may also influence ligament stretching. For D–P collisions at the same $We_d$, a TGB particle has higher inertia, corresponding to a larger $We_p$. This suggests that the combined D–P system after collision possesses greater rotational energy. The resulting centrifugal forces could enhance ligament extension and thinning, accelerating the separation process. This could explain why the separation time is similar for TGB and PB in Fig. \ref{separation}.

After the particle pinches off from the droplet, the ligament disintegrates and contracts into a small secondary droplet. Previous studies have shown that the number of secondary droplets formed during D–P collisions depends on the impact parameter, $B$ \cite{pawar2016experimental}. However, across the range of impact conditions considered in this study, only a single secondary droplet was observed during collisions of the GB and TGB particles with the water droplet. In case of PB, no secondary droplet formation was recorded. The formation of secondary droplets may depend on the combined effects of $\theta_w$, $\lambda$, and viscous-capillary interactions represented by the Ohnesorge number, $Oh$. A systematic study investigating the effects of these factors on secondary droplet formation presents an interesting direction for future research.

Overall, the analysis of experimental images in Fig. \ref{DPlow} and Fig. \ref{separation} shows that the forces acting on the particle and at the droplet–particle (D-P) interface play a central role in determining the collision outcome. In contrast to $We_d$, the particle Weber number, $We_p$, provides a more appropriate scaling for these interactions, as it accounts for the particle inertia relative to the restoring capillary forces at the D–P interface. As a consequence, $We_p$ serves as a unifying parameter that facilitates clearer interpretation of the observed collision behaviour at $\Delta=3$. This is further evidenced by the similar collision outcomes observed for high-offset collisions involving a water droplet and a PB particle, and a water droplet and a TGB particle, at fixed $We_p$ (\textcolor{blue}{refer to S4 in the Supplementary Material}).

To investigate the influence of $We_p$ on collision outcomes across different $B$, regime maps were constructed to delineate transitions between particle capture and separation. Figures \ref{reg_map}a, \ref{reg_map}b, and \ref{reg_map}c depict the capture (black symbols) and separation (red symbols) outcomes for water droplets colliding with GB, TGB, and PB particles, respectively. Figure \ref{reg_map}d illustrates the regime boundary lines separating the collision outcomes for all three particles. As evident, the maximum $We_p$ for PB particles is lower than that of TGB and GB particles due to lower density.

For all three systems, at low offset $B$, particles were captured across the entire range of $We_p$ tested. This outcome is consistent with previous experimental studies on moving D–P systems by Le Gac \& Planchette \cite{le2024novel} and Pawar et al. \cite{pawar2016experimental}. Hereafter, these studies are referred to as Le Gac and Pawar, respectively. A distinct behaviour was observed for GB, a strong wetting particle. At high $B$, the particle separated from the droplet when $We_p$ was sufficiently high. However, at low $v_{rel}$, corresponding to lower particle Weber numbers ($We_p < 30$), particle capture occurred even for glancing collisions ($B \approx 1)$, representing a novel observation of the present study. This behavior can be attributed to the strong capillary forces exerted by the droplet on the hydrophilic GB. Pawar and Le Gac did not observe this effect, likely because their studies employed partially wettable particles ($\theta_w \approx 75\degree$).

In contrast, both PB and TGB beads, with comparable hydrophobicity ($\theta_w = 90\degree$ and $\theta_w = 88\degree$ respectively), exhibited separation outcomes during glancing collisions throughout the entire range of $We_p$ considered in this study. This observation aligns with trends reported in the literature. For PB, the experimental range of $We_p$ became narrower due to a lower $\lambda$ compared to GB and TGB. For TGB, the maximum and minimum experimental $We_p$ shifted to values above 10, which can be attributed to its higher $\lambda$.

Overall, $We_p$ effectively captures the influence of particle inertia. This is evident in Fig. \ref{reg_map}d, where the regime boundaries for TGB and PB coincide despite their differing $\rho_p$. Also, both TGB and PB regime boundaries exhibit a steep decline up to $We_p \approx 20$, marking the upper experimental limit for PB. Beyond $We_p \approx 30$, the TGB regime boundary plateaus. As discussed earlier, experimental images for TGB show a transition in D–P interaction at $B \approx 0$ with increasing $We_d$ (Fig. \ref{DPlow}), where the particle penetrates the droplet rather than being captured at the interface. For a particle captured at the droplet interface, its inertia must overcome the capillary forces acting at the interface to achieve separation from the droplet. In contrast, a penetrated particle must overcome both the capillary forces and the viscous drag within the droplet to detach.

A transition in D–P interaction from capture to penetration can also occur at higher $B$. Although the present experimental images do not provide definitive confirmation, we believe that the TGB particle begins to penetrate the droplet when $We_p > 30$ and $B \approx 0.6$. Beyond this limit, the TGB particle experiences combined capillary and viscous resistance from the droplet, which increases with decreasing $B$ due to the larger droplet–particle interaction area at lower $B$ values. Moreover, the viscous drag also increases with increasing $v_r$ and, therefore, $We_p$. These cumulative effects could explain why the TGB regime boundary plateaus beyond $We_p > 30$ in Fig. \ref{reg_map}d. Future experiments employing transparent droplets, achieved through illumination adjustments, could help determine the precise threshold for particle penetration at different $B$ during D–P collisions and provide deeper insight into the underlying physics.

However, the definition of $We_p$ assumes that the change in interfacial area generated during collision scales as $\sim D_p^2$. This neglects the work done by surface tension in unwrapping the liquid from the particle surface during separation, which depends on wettability \cite{zhu2025dynamics}. All three particles, GB, TGB, and PB, have similar $D_p$, but the wettability-dependent energy barrier is higher for the hydrophilic GB particle. This explains why the GB regime boundary lies above those of TGB and PB.

In addition, moving D–P collisions must account for both droplet and particle inertia when modeling the collision process and characterizing the outcomes. This motivates the introduction of a new effective particle Weber number, $We_{peffw}$, which is derived and discussed later in this section. In the present study, D–P collisions were performed at fixed $\Delta$. The merging of the TGB and PB regime boundaries in Fig.~\ref{reg_map} indicates that droplet inertia, while significant, remains comparable across all cases examined. This enables $We_p$ to serve as a consistent parameter for comparing collision outcomes.

We further compared the current experimental results with the findings reported by Pawar \cite{pawar2016experimental}, Le Gac \cite{le2024novel}, and Al-Dirawi et al. \cite{al2020experimental}(hereafter referred to as Al-Dirawi). Both Le Gac \cite{le2024novel} and Pawar \cite{pawar2016experimental} investigated D–P collisions involving moving droplets and particles. In Le Gac’s study, the size and density ratios were fixed at $\Delta = 1$ and $\lambda = 1$, respectively, and the test particles were partially wetting, with an equilibrium contact angle of $\theta_w = 70\degree$. The droplets employed were viscous, with $\mu = 0.0049$ Pa·s ($Oh_d = 0.031$). Pawar also used partially wetting particles ($\theta_w = 75^\circ$), which were impacted by non-viscous distilled water droplets under conditions $\Delta = 1.17$, $\lambda = 2.5$, and $Oh_d = 0.0022$. Al-Dirawi conducted droplet–droplet (D–D) collisions at $\Delta = 1$ and $\lambda = 1$ using droplets of non-identical viscosity. The droplets were prepared with 2\% and 8\% HPMC solutions, giving viscosity values of $\mu = 0.0028$ Pa·s and $\mu = 0.0284$ Pa·s, respectively, thereby making one droplet significantly more resistant to deformation. The $Oh_d$ value associated with the less viscous droplet was 0.021.

Both $\Delta$ and $\lambda$ influence the momentum of the droplet and particle prior to collision and, as noted, differ between the literature and the present study. To enable a consistent evaluation of impact conditions, an effective dimensionless parameter must be defined that accounts for the dynamics of both the droplet and particle.

Consider a head-on collision scenario between a droplet and a particle. In a zero-momentum frame of reference, the reference velocity can be evaluated as:

\begin{equation}
    v_{ref} = \frac{m_d v_d - m_p v_p}{m_p + m_d}
\end{equation}
Here, $m_d$ and $m_p$ are the mass of the droplet and the particle respectively. Subsequently, the effective droplet velocity, $v_{deff}$, and the effective particle velocity, $v_{peff}$ (in zero-momentum frame of reference) can be written as: 
\begin{equation}
    v_{deff} = v_d - v_{ref} = \frac{m_p (v_d + v_p)}{m_p + m_d}
\end{equation}
\begin{equation}
    v_{peff} = v_p + v_{ref} = \frac{m_d (v_d + v_p)}{m_p + m_d} 
\end{equation}
Therefore, the kinetic energy of the combined D-P system can be expressed as: 
\begin{equation}
    KE_{eff} \sim \frac{ \rho_p D_p^3 v_r^2 \Delta^3}{\lambda+\Delta^3}.
    \label{Kereff}
\end{equation}
When the droplet and particle have equal size and density, Eq. \ref{Kereff} reduces to one-half of the relative kinetic energy. Thus, only half of the available energy is available for deformation and dissipation during the collision, while the remainder is retained as translational kinetic energy of the center of mass. In contrast, when one body is stationary, the entirety of $KE_{eff}$ contributes to the collision.

\begin{figure*}[ht]
	\centering
		\includegraphics[scale=0.6]{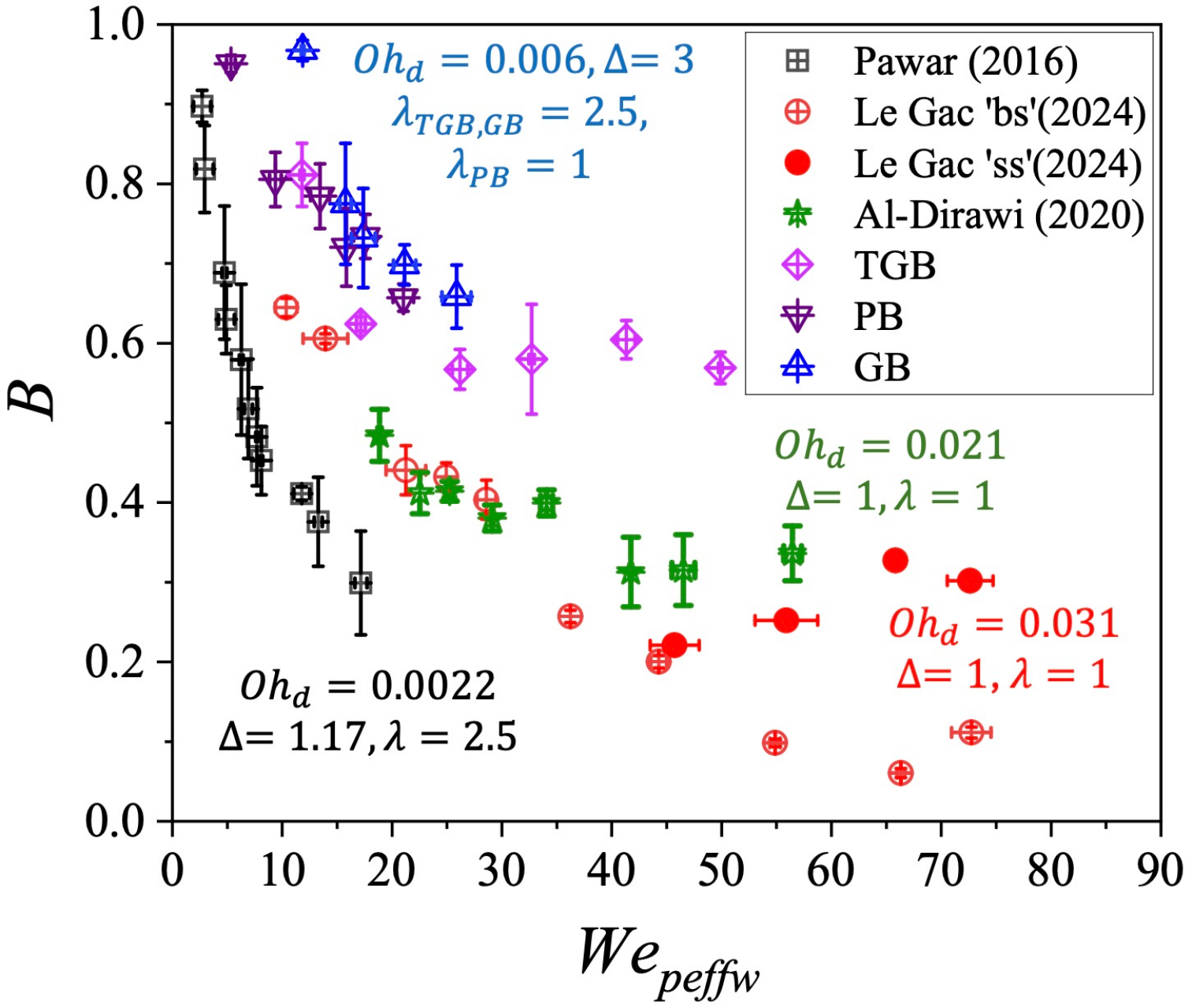}
	\caption{ Regime boundaries separating particle capture and separation outcomes, plotted as $We_{peffw}$ versus $B$ for the present experiments on GB, TGB, and PB, and compared with data from the literature.}
	\label{lit_comp}
\end{figure*}

In the inviscid limit, separation between the droplet and particle after collision can occur only if $KE_{eff}$ exceeds the change in droplet surface energy (before and after particle detachment), $\Delta SE$, as well as the work done by surface tension in unwrapping the droplet from the particle, $W_{\theta}$ \cite{zhu2025dynamics}. As droplet–particle (D–P) collisions are transient in nature, accurate estimation of $\Delta SE$ and $W_{\theta}$ requires knowledge of the instantaneous droplet shape and the evolving contact angle at the TPCL. Here, we assume that $\Delta SE$ scales as $\sim \sigma D_p^2$, while $W_{\theta}$ scales as $\sim \sigma D_p^2 \cos\theta_w$. Taking a ratio of $KE_{eff}$ and $\Delta SE + W_{\theta}$ yields:

\begin{equation}
   \frac{KE_{eff}}{SE + W_{\theta}} \sim We_p \left( \frac{\Delta^3}{\lambda + \Delta^3} \right) \left( \frac{1}{1 + cos \theta_w} \right) = We_{peffw}
\label{Wepeffw}
\end{equation}

We define the new dimensionless term in Eq.~\ref{Wepeffw} as the effective particle Weber number, $We_{peffw}$. In the limiting cases $\Delta \to 0$ and $\Delta \to \infty$, Eq.~\ref{Wepeffw} simplifies to $We_d$ and $We_p$, respectively (\textcolor{blue}{refer to S5 in the Supplementary Material}). In addition, if $W_\theta$ is neglected, Eq.~\ref{Wepeffw} reduces to $\frac{KE_{eff}}{SE} \sim We_p\left( \frac{\Delta^3}{\lambda + \Delta^3} \right)$. We term this dimensionless number as $We_{peff}$. For the present experimental dataset, substituting the values of $\Delta$ and $\lambda$ yields $We_{peff} \approx 0.9 We_p$ for collision data of GB and TGB particles, and $We_{peff} \approx 0.96 We_p$ for the collision data of PB particles. In other words, for the D–P collisions investigated in the present study, at least $90\%$ of $KE_{eff}$ contributes to the collision. This effect is evident in the $B$ vs $We_p$ plot and the $B$ vs $We_{peff}$ inset in Fig.~\ref{reg_map}d. A comparison of these plots shows negligible differences in the observed trends of the regime boundary lines for the GB, TGB, and PB particles investigated in this study.

Figure \ref{lit_comp} compares the regime boundaries separating D–P collision outcomes: particle capture and separation, observed in the present experiments with those reported in the literature, plotted in terms of $We_{peffw}$ versus $B$. 

In general, all regime boundary lines show similar qualitative behaviour: at high $B$, a small $We_{peffw}$ is sufficient to induce separation, whereas at low $B$, progressively larger $We_{peffw}$ is required. Despite this similarity, clear differences in trend are evident across the datasets. At a given $B$, the critical $We_{peffw}$ required for separation is lowest in Pawar's experiments. This is followed by the critical $We_{peffw}$ exhibited in the datasets of Le Gac and Al-Dirawi, and then by the present experiments on GB, TGB, and PB, which require the highest $We_{peffw}$ values to separate from the droplet. 

During droplet–particle (D–P) collisions, the droplet resists deformation caused by particle inertia through viscous and capillary forces. If these resisting forces are insufficient, the particle inertia can deform the droplet beyond a critical limit, leading to separation or fragmentation; otherwise, the particle is captured.
The droplet’s resistance to deformation, or its stability, is quantified by the droplet Ohnesorge number, $Oh_d$. In Pawar’s experiments, $Oh_d$ is very small ($Oh_d = 0.0022$), indicating that viscous resistance is negligible and the separation outcome is primarily governed by the balance between inertial and capillary forces. Although the droplet-to-particle size ratio, $\Delta$, in Pawar’s experiments is similar to that in the studies of Le Gac and Al-Dirawi, the $Oh_d$ values in the latter are an order of magnitude higher ($Oh_d = 0.031$ and $0.021$, respectively). Higher $Oh_d$ increases the droplet’s resistance to deformation, which suppresses particle separation. This explains why the critical $We_{peffw}$ required for separation in the datasets of Le Gac and Al-Dirawi is higher than in Pawar’s experiments.

An intriguing phenomenon emerges when comparing the regime boundaries of Le Gac and Al-Dirawi. With similar  $\Delta$ and $\lambda$, and close values of $Oh_d$, the boundaries appear to overlap until $We_{peffw}\approx 30$, beyond which the Le-Gac boundary splits into two. The Le-Gac ‘ss’ boundary (closed red symbols) denotes the transition between stretching and bag separation, while the Le-Gac ‘bs’ boundary (open red symbols) marks the transition between bag separation and particle capture (termed “deposition” by Le Gac \cite{le2024novel}). Bag separation is characterised by the formation of a “bag” shape of the droplet upon collision with a solid particle, typically at low $B$. The bag subsequently recoils and breaks up into smaller droplets. The Al-Dirawi boundary separates stretch-separation collision outcomes from coalescence, observed in the case of D-D collision with non-identical viscosity. As shown in Fig. \ref{lit_comp}, for $We_{peffw} > 30$, the Le Gac ‘bs’ boundary continues to decline, while the Le Gac ‘ss’ and Al-Dirawi boundaries remain close to each other and plateau. This transformation of the regime boundary line reflects a shift in the governing force balance as the separation mechanism changes.

In the preceding section, we discussed the change in the shape of the regime boundary observed for TGB in the present experiments (Fig. \ref{reg_map}d). The boundary exhibits a plateau when $We_p > 30$ and $B \approx 0.6$. Beyond this limit, the particle begins to penetrate the droplet. This alters the energy balance, as the particle is decelerated by capillary forces at the interface and by viscous dissipation within the droplet. A similar behaviour is observed for TGB in Fig. \ref{lit_comp} when $We_{peffw} > 20$.

Another reason for the plateauing may be viscous dissipation associated with the region of interaction between the droplet and the particle. As $B$ decreases, the overlap between the droplet and particle increases. This leads to a larger interaction region and enhanced viscous dissipation. Both the Le Gac ‘ss’ and Al-Dirawi regime boundaries, which correspond to stretching separation, also plateau when $We_{peffw} > 30$ and $B \approx 0.3$, and remain close to each other.

Since Al-Dirawi conducted D–D collision experiments, the plateauing of their regime boundary can be attributed to increased viscous dissipation associated with the larger region of interaction at low $B$. The simultaneous plateauing observed for the Le Gac ‘ss’ regime boundary indicates that both regime boundaries are governed by the same underlying physics.

An additional factor that may contribute to the plateauing of the Le Gac ‘ss’ and Al-Dirawi regime boundaries is the formation of a thicker ligament between the droplet and particle, or between the droplets, at smaller $B$. The larger overlap leads to a thicker ligament, which stretches more before breakup. This increases the energy barrier for separation and suppresses particle detachment, resulting in a plateau of the regime boundary.

Notably, the Le Gac ‘ss’ and Al-Dirawi regime boundaries remain close to each other. This suggests that the force balance governing stretching separation in D–D collisions is similar to that in D–P collisions. This similarity may arise because a more viscous droplet resists deformation in a manner comparable to a solid particle.

Compared to the literature, the regime boundaries for GB, TGB, and PB in the present experiments exhibit the highest critical $We_{peffw}$ required for separation. This increased threshold, despite a relatively small $Oh_d$ of 0.006, suggests that the larger $\Delta$, which is equal to 3 in the present experiments and the highest among the reported studies, influences the impact geometry and enhances the resistance to separation.

The underlying physics can be illustrated through a simple scaling analysis. Consider a D–P collision scenario. The viscous resistance experienced by the particle after impact can be approximated as
$E_{diss} \sim \phi V_{diss} t_c$ \cite{chandra1991collision}. Here, the viscous dissipation function is $\phi \sim \mu (v_r / D_p)^2$,
the dissipated volume is estimated as $V_{diss} \sim D_p^2 D_d$, and the characteristic timescale corresponds to the advection time, $t_c \sim D_d / v_r$. Substituting these expressions gives the following estimate for the dissipated energy:

\begin{equation}
    E_{diss} \sim \mu v_r D_d^2 
\end{equation}

Taking a ratio of $E_{diss}$ and $KE_{eff}$ gives:

\begin{equation}
    E_{diss}/KE_{eff} \sim  \frac{Oh_d \Delta^{2.5}}{\sqrt{We_{peff}\lambda}}  
    \label{Energyratio}
\end{equation}

Equation \ref{Energyratio} shows that, at a given $We_{peff}$, the fraction of energy lost to viscous dissipation increases with $Oh_d$ and follows a power-law scaling with the size ratio, $\Delta^{2.5}$. This indicates that, for a fixed particle size, larger droplets involve a greater volume of fluid being sheared during the collision, which enhances viscous resistance to deformation.

In the present experiments, the D–P systems (GB, TGB, and PB) have larger $\Delta$ than those reported in the literature. This implies stronger viscous resistance during collision, which explains why a higher critical $We_{peffw}$ is required for separation, even though $Oh_d$ is smaller than in the experiments of Le Gac and Al-Dirawi.

Equation~\ref{Energyratio} also indicates that $\lambda$ influences the relative energy lost to viscous dissipation by affecting particle deceleration following the droplet–particle interaction. However, the effect of $\lambda$ is less pronounced at low $We_{peffw}$. As a result, for the same $\Delta$, the regime boundaries of GB, TGB, and PB in the present experiments converge at low $We_{peffw}$ ($We_{peffw} \leq 30$).

Nonetheless, Eq. \ref{Energyratio} is derived from scaling arguments and requires validation. Extending experiments covering a wider range of $\Delta$ would not only test this scaling but also provide deeper insight into the role of droplet size in D–P collision dynamics, representing a promising direction for future research.

Overall, the analysis of Fig. \ref{lit_comp} shows that the outcomes of D–P collisions depend on droplet and particle properties, impact conditions, the characteristic length scale of the system, and the dominant interaction mechanism. The modified Weber number $We_{peffw}$, proposed in the present work, successfully captures the role of particle properties, as indicated by the convergence of regime boundary points for GB, TGB, and PB beads. However, $We_{peffw}$ does not account for the viscous resistance exerted by the droplet on the particle. Although $\Delta$ is embedded in the definition of $We_{peffw}$, it also governs the geometry of the D–P collision. Hence, its influence on viscous resistance remains implicit and is not captured in the present formulation. Incorporating $Oh_d$ alongside $We_{peffw}$ could provide a framework that includes both viscosity and droplet length scale. Even so, the explicit role of $\Delta$ on the boundaries would remain unresolved. Addressing this limitation presents a central challenge for future modeling efforts. The partial overlap between the Le Gac and Karrar regime boundaries in Fig. \ref{lit_comp} also suggests a possible route to simplify free droplet–particle experiments by substituting them with droplet–droplet collisions of non-identical viscosity. Although D–D collisions cannot capture the effects of particle wettability, they can reveal the influence of geometry. Hence, D–D collisions with non-identical viscosity present a promising direction for future research, provided systematic validation is carried out across a wider range of $\Delta$.

Establishing a predictive model for the critical regime boundary in droplet–particle (D–P) collisions requires a systematic assessment of energy redistribution before and after impact over a range of $\Delta$ and $B$. This requires knowledge of the instantaneous droplet shape and the dynamic contact angle at the TPCL, which can be obtained through targeted experiments and high-fidelity simulations.

An alternative approach is critical ligament length analysis, which has predicted regime transitions in droplet–droplet collisions \cite{al2021inertial}. However, in D–P collisions, the ligament length depends on the particle contact angle. As a result, this approach cannot be applied without modification. This highlights the need for a framework that incorporates wettability effects in ligament-based criteria.

\section{Conclusions}
The study investigates mid-air collisions between free micron-sized spherical droplets and particles with a droplet-to-particle size ratio, $\Delta \approx 3$, at varying collision offset, $B$. The effects of particle density, $\rho_p$, and wettability, $\theta_w$, on the collision outcomes are examined using three representative particles (GB, TGB, and PB). The results demonstrate that $\rho_p$ governs whether a particle is engulfed or remains at the droplet interface after impact, while low $\theta_w$ suppresses particle separation even during glancing collisions. The conventional droplet Weber number is shown to be inadequate for mapping collision outcomes, as it does not account for particle inertia at large size ratios.

By formulating the collision dynamics in a zero-momentum frame of reference, a new effective Weber number, $We_{peffw}$, is introduced to incorporate particle inertia and wettability in characterizing collision regimes. Regime maps constructed in the $We_{peffw}$ versus $B$ space exhibit collapse of regime boundaries for the present data. When combined with literature results in a unified regime map, the collapse or the presence of distinct trends in the regime boundaries is found to depend on the Ohnesorge number, $Oh_d$, which governs viscous resistance to deformation, and the size ratio, $\Delta$, which controls collision geometry.

The results also indicate that droplet–droplet (D–D) collisions involving droplets of non-identical viscosity can reproduce key features of droplet–particle (D–P) interactions, providing a more accessible experimental route to probe collision dynamics. 

The findings of the present study on droplet–particle (D-P) collisions could have significant impact on the development of predictive models to control agglomeration in spray drying and particle capture in aerosol scavenging. 

\section{Acknowledgements}
This work was supported by the EPSRC project ‘Evaporative Drying of Droplets and the
Formation of Micro-structured and Functional Particles and Films’ (grant no. EP/N025245/1), UK, and EPSRC-UK Future Manufacturing Hub in Manufacture using Advanced Powder Processes (MAPP) (grant no. EP/P006566/1). 

\section*{Data availability}
The data that support the findings of this study are available from the corresponding author upon reasonable request.

\section{References}
\bibliography{aipsamp}

@article{al2019new,
  title={A new model for the bouncing regime boundary in binary droplet collisions},
  author={Al-Dirawi, Karrar H and Bayly, Andrew E},
  journal={Physics of fluids},
  volume={31},
  number={2},
  year={2019},
  publisher={AIP Publishing}
}

@article{dubrovsky1992particle,
  title={Particle interaction in three-phase polydisperse flows},
  author={Dubrovsky, VV and Podvysotsky, AM and Shraiber, AA},
  journal={International journal of multiphase flow},
  volume={18},
  number={3},
  pages={337--352},
  year={1992},
  publisher={Elsevier}
}

@article{le2024novel,
  title={A novel experimental approach to study drop-particle collisions},
  author={Le Gac, Jean-Baptiste and Planchette, Carole},
  journal={Atomization and Sprays},
  volume={34},
  number={4},
  year={2024},
  publisher={Begel House Inc.}
}

@article{pawar2016experimental,
  title={An experimental study of droplet-particle collisions},
  author={Pawar, Sandip K and Henrikson, Filip and Finotello, Giulia and Padding, Johan T and Deen, Niels G and Jongsma, Alfred and Innings, Fredrik and Kuipers, JAM Hans},
  journal={Powder Technology},
  volume={300},
  pages={157--163},
  year={2016},
  publisher={Elsevier}
}

@article{sechenyh2016experimental,
  title={An experimental study for impact of a drop onto a particle in mid-air: The influence of particle wettability},
  author={Sechenyh, Vitaliy and Amirfazli, Alidad},
  journal={Journal of Fluids and Structures},
  volume={66},
  pages={282--292},
  year={2016},
  publisher={Elsevier}
}

@article{wu2016abatement,
  title={Abatement of fine particle emission by heterogeneous vapor condensation during wet limestone-gypsum flue gas desulfurization},
  author={Wu, Hao and Pan, Danping and Huang, Rongting and Hong, Guangxin and Yang, Bing and Peng, Ziming and Yang, Linjun},
  journal={Energy \& Fuels},
  volume={30},
  number={7},
  pages={6103--6109},
  year={2016},
  publisher={ACS Publications}
}

@article{malgarinos2017numerical,
  title={Numerical investigation of heavy fuel droplet-particle collisions in the injection zone of a Fluid Catalytic Cracking reactor, Part I: Numerical model and 2D simulations},
  author={Malgarinos, Ilias and Nikolopoulos, Nikolaos and Gavaises, Manolis},
  journal={Fuel Processing Technology},
  volume={156},
  pages={317--330},
  year={2017},
  publisher={Elsevier}
}

@article{byrne1993scavenging,
  title={Scavenging of sub-micrometre aerosol particles by water drops},
  author={Byrne, MA and Jennings, SG},
  journal={Atmospheric Environment. Part A. General Topics},
  volume={27},
  number={14},
  pages={2099--2105},
  year={1993},
  publisher={Elsevier}
}

@article{wu1992interaction,
  title={Interaction mechanisms between ceramic particles and atomized metallic droplets},
  author={Wu, Yue and Lavernia, Enrique J},
  journal={Metallurgical Transactions A},
  volume={23},
  pages={2923--2937},
  year={1992},
  publisher={Springer}
}

@article{tinsley2000effects,
  title={Effects of image charges on the scavenging of aerosol particles by cloud droplets and on droplet charging and possible ice nucleation processes},
  author={Tinsley, BA and Rohrbaugh, RP and Hei, M and Beard, KV},
  journal={Journal of the atmospheric sciences},
  volume={57},
  number={13},
  pages={2118--2134},
  year={2000}
}

@article{ardon2015laboratory,
  title={Laboratory studies of collection efficiency of sub-micrometer aerosol particles by cloud droplets on a single-droplet basis},
  author={Ardon-Dryer, Karin and Huang, Y-W and Cziczo, Daniel J},
  journal={Atmospheric Chemistry and Physics},
  volume={15},
  number={16},
  pages={9159--9171},
  year={2015},
  publisher={Copernicus GmbH G{\"o}ttingen, Germany}
}

@article{zhan2023droplet,
  title={Droplet-particle collision dynamics: A molecular dynamics simulation},
  author={Zhan, Lingxiao and Chen, Heng and Zhou, Hao and Chen, Jiawei and Wu, Hao and Yang, Linjun},
  journal={Powder Technology},
  volume={422},
  pages={118456},
  year={2023},
  publisher={Elsevier}
}

@article{tkachenko2022experimental,
  title={Experimental research of liquid droplets colliding with solid particles in a gaseous medium},
  author={Tkachenko, Pavel P and Shlegel, NE and Strizhak, Pavel A},
  journal={Chemical Engineering Research and Design},
  volume={177},
  pages={200--209},
  year={2022},
  publisher={Elsevier}
}

@article{Tenou,
title = {Batch and continuous fluid bed coating  review and state of the art},
journal = {Journal of Food Engineering},
volume = {53},
number = {4},
pages = {325-340},
year = {2002},
issn = {0260-8774},
author = {E Teunou and D Poncelet},
}

@article{yoon2022adaptive,
  title={Adaptive mesh axi-symmetric simulation of droplet impact with a spherical particle in mid-air},
  author={Yoon, Ikroh and Chergui, Jalel and Juric, Damir and Shin, Seungwon},
  journal={International Journal of Multiphase Flow},
  volume={155},
  pages={104193},
  year={2022},
  publisher={Elsevier}
}

@article{mitra2015collision,
  title={Collision behaviour of a smaller particle into a larger stationary droplet},
  author={Mitra, Subhasish and Doroodchi, Elham and Pareek, Vishnu and Joshi, Jyeshtharaj B and Evans, Geoffrey M},
  journal={Advanced Powder Technology},
  volume={26},
  number={1},
  pages={280--295},
  year={2015},
  publisher={Elsevier}
}

@article{geng2023collision,
  title={Collision regimes and dynamic behaviors of a viscous droplet impacting on a spherical particle at high temperatures},
  author={Geng, Pengfei and Ma, Jiliang and Chen, Xiaoping and Liu, Daoyin and Pan, Suyang and Liang, Cai},
  journal={Physics of Fluids},
  volume={35},
  number={3},
  year={2023},
  publisher={AIP Publishing}
}

@article{banitabaei2017droplet,
  title={Droplet impact onto a solid sphere: Effect of wettability and impact velocity},
  author={Banitabaei, SA and Amirfazli, A},
  journal={Physics of Fluids},
  volume={29},
  number={6},
  year={2017},
  publisher={AIP Publishing}
}

@article{islamova2022droplet,
  title={Droplet-droplet, droplet-particle, and droplet-substrate collision behavior},
  author={Islamova, AG and Kerimbekova, SA and Shlegel, NE and Strizhak, PA},
  journal={Powder Technology},
  volume={403},
  pages={117371},
  year={2022},
  publisher={Elsevier}
}

@article{speirs2023capture,
  title={The capture of airborne particulates by rain},
  author={Speirs, Nathan B and Belden, Jesse L and Hellum, Aren M},
  journal={Journal of Fluid Mechanics},
  volume={958},
  pages={A40},
  year={2023},
  publisher={Cambridge University Press}
}

@article{wu2021simulating,
  title={Simulating the collision of a moving droplet against a moving particle: Impact of Bond number, wettability, size ratio, and eccentricity},
  author={Wu, Guoqiang and Chen, Sheng},
  journal={Physics of Fluids},
  volume={33},
  number={9},
  year={2021},
  publisher={AIP Publishing}
}

@article{frohlich2023nozzle,
  title={Nozzle zone agglomeration in spray dryers: Determination of the agglomeration efficiency in the fines return by means of agglomerate properties and residence time distribution},
  author={Fr{\"o}hlich, Jakob Alfons and Ruprecht, Nora Alina and Kohlus, Reinhard},
  journal={Drying Technology},
  volume={41},
  number={12},
  pages={1907--1923},
  year={2023},
  publisher={Taylor \& Francis}
}

@article{al2020experimental,
  title={An experimental study of binary collisions of miscible droplets with non-identical viscosities},
  author={Al-Dirawi, Karrar H and Bayly, Andrew E},
  journal={Experiments in Fluids},
  volume={61},
  pages={1--22},
  year={2020},
  publisher={Springer}
}

@article{al2021inertial,
  title={Inertial stretching separation in binary droplet collisions},
  author={Al-Dirawi, Karrar H and Al-Ghaithi, Khaled HA and Sykes, Thomas C and Castrej{\'o}n-Pita, J Rafael and Bayly, Andrew E},
  journal={Journal of Fluid Mechanics},
  volume={927},
  pages={A9},
  year={2021},
  publisher={Cambridge University Press}
}

@article{yang2018simulation,
  title={Simulation of interaction between a freely moving solid particle and a freely moving liquid droplet by lattice Boltzmann method},
  author={Yang, Bo and Chen, Sheng},
  journal={International Journal of Heat and Mass Transfer},
  volume={127},
  pages={474--484},
  year={2018},
  publisher={Elsevier}
}

@article{chen2018entrapping,
  title={Entrapping an impacting particle at a liquid--gas interface},
  author={Chen, Han and Liu, Hao-Ran and Lu, Xi-Yun and Ding, Hang},
  journal={Journal of Fluid Mechanics},
  volume={841},
  pages={1073--1084},
  year={2018},
  publisher={Cambridge University Press}
}

@article{lee2008impact,
  title={Impact of a superhydrophobic sphere onto water},
  author={Lee, Duck-Gyu and Kim, Ho-Young},
  journal={Langmuir},
  volume={24},
  number={1},
  pages={142--145},
  year={2008},
  publisher={ACS Publications}
}

@article{vella2015floating,
  title={Floating versus sinking},
  author={Vella, Dominic},
  journal={Annual Review of Fluid Mechanics},
  volume={47},
  number={1},
  pages={115--135},
  year={2015},
  publisher={Annual Reviews}
}

@article{chandra1991collision,
  title={On the collision of a droplet with a solid surface},
  author={Chandra, S and Avedisian, CT},
  journal={Proceedings of the Royal Society of London. Series A: Mathematical and Physical Sciences},
  volume={432},
  number={1884},
  pages={13--41},
  year={1991},
  publisher={The Royal Society London}
}

@article{zhu2025dynamics,
  title={Dynamics of rebound in head-on collisions between suspended micrometer-sized droplets and particles},
  author={Zhu, Xiaolong and Xia, Pengzhi and Pan, Chuanyu and Wang, Wenjie and Tao, Ruiqing and Zhang, Jinyi and Han, Fangwei and Xu, Chaohang and Long, Biaoli and Yang, Husheng and others},
  journal={Physics of Fluids},
  volume={37},
  number={4},
  year={2025},
  publisher={AIP Publishing}
}

@article{lobel2021interparticle,
  title={Interparticle Repulsion of Microparticles Delivered to a Pendent Drop by an Electric Field},
  author={Lobel, Benjamin T and Hobson, Matthew J and Ireland, Peter M and Webber, Grant B and Thomas, Casey A and Ogino, Haruka and Fujii, Syuji and Wanless, Erica J},
  journal={Langmuir},
  volume={38},
  number={2},
  pages={670--679},
  year={2021},
  publisher={ACS Publications}
}

@article{yoon2021maximal,
  title={Maximal spreading of droplet during collision on particle: Effects of liquid viscosity and surface curvature},
  author={Yoon, Ikroh and Shin, Seungwon},
  journal={Physics of Fluids},
  volume={33},
  number={8},
  year={2021},
  publisher={AIP Publishing}
}

@article{hardalupas1999experimental,
  title={Experimental investigation of sub-millimetre droplet impingement on to spherical surfaces},
  author={Hardalupas, Y and Taylor, AMKP and Wilkins, JH},
  journal={International journal of heat and fluid flow},
  volume={20},
  number={5},
  pages={477--485},
  year={1999},
  publisher={Elsevier}
}

@article{bakshi2007investigations,
  title={Investigations on the impact of a drop onto a small spherical target},
  author={Bakshi, Shamit and Roisman, Ilia V and Tropea, Cam},
  journal={Physics of fluids},
  volume={19},
  number={3},
  year={2007},
  publisher={AIP Publishing}
}

@article{charalampous2017collisions,
  title={Collisions of droplets on spherical particles},
  author={Charalampous, Georgios and Hardalupas, Yannis},
  journal={Physics of Fluids},
  volume={29},
  number={10},
  year={2017},
  publisher={AIP Publishing}
}

@article{khurana2019phenomenology,
  title={Phenomenology of droplet collision hydrodynamics on wetting and non-wetting spheres},
  author={Khurana, Gargi and Sahoo, Nilamani and Dhar, Purbarun},
  journal={Physics of Fluids},
  volume={31},
  number={7},
  year={2019},
  publisher={AIP Publishing}
}

@article{mitra2013droplet,
  title={Droplet impact dynamics on a spherical particle},
  author={Mitra, Subhasish and Sathe, Mayur J and Doroodchi, Elham and Utikar, Ranjeet and Shah, Milin K and Pareek, Vishnu and Joshi, Jyeshtharaj B and Evans, Geoffrey M},
  journal={Chemical Engineering Science},
  volume={100},
  pages={105--119},
  year={2013},
  publisher={Elsevier}
}

@article{bordbar2018maximum,
  title={Maximum spreading and rebound of a droplet impacting onto a spherical surface at low Weber numbers},
  author={Bordbar, Alireza and Taassob, Arsalan and Khojasteh, Danial and Marengo, Marco and Kamali, Reza},
  journal={Langmuir},
  volume={34},
  number={17},
  pages={5149--5158},
  year={2018},
  publisher={ACS Publications}
}

@article{yoon2022promoting,
  title={Promoting rebound from droplet impact on a spherical particle: Experimental and numerical study},
  author={Yoon, Ikroh and Ha, Chiwook and Lee, Choongyeop and Shin, Seungwon},
  journal={Physics of Fluids},
  volume={34},
  number={10},
  year={2022},
  publisher={AIP Publishing}
}

@article{khojasteh2019review,
  title={A review of liquid droplet impacting onto solid spherical particles: A physical pathway to encapsulation mechanisms},
  author={Khojasteh, Danial and Kazerooni, Nooshin Moradi and Marengo, Marco},
  journal={Journal of industrial and engineering chemistry},
  volume={71},
  pages={50--64},
  year={2019},
  publisher={Elsevier}
}

@article{sykes2022droplet,
  title={Droplet splashing on curved substrates},
  author={Sykes, Thomas C and Fudge, Ben D and Quetzeri-Santiago, Miguel A and Castrej{\'o}n-Pita, J Rafael and Castrej{\'o}n-Pita, Alfonso A},
  journal={Journal of Colloid and Interface Science},
  volume={615},
  pages={227--235},
  year={2022},
  publisher={Elsevier}
}

@article{fan2024numerical,
  title={Numerical simulation of mid-air collisions between droplets and particles: An examination of particle forces and kinetic energy dissipation},
  author={Fan, Zhiheng and Liu, Daoyin and Liang, Cai and Chen, Xiaoping},
  journal={Powder Technology},
  volume={432},
  pages={119124},
  year={2024},
  publisher={Elsevier}
}

\end{document}